# EnergyAgentBench: Benchmarking LLM Agents on Live Energy Infrastructure Data

Eliseo Curcio

**Abstract**

Selecting the right electricity market region for a hyperscale AI datacenter requires reasoning across live electricity prices, grid carbon intensity, technology cost trajectories, and causal grid dynamics - a multi-step, multi-source analytical task that static knowledge benchmarks cannot evaluate. We introduce EnergyAgentBench, the first agentic benchmark grounded in live electricity market data for this problem class. The benchmark comprises 70 task variants across five families: datacenter siting under cost-carbon trade-offs (F1), long-horizon portfolio siting (F1-LH), lifetime LCOE ranking over multi-decade cost trajectories (F2), 30-year portfolio optimization (F2-LH), and causal grid diagnosis (F3). Tasks require 3 to 48 sequential tool calls against live endpoints from the QuarluxAI infrastructure platform, the U.S. Energy Information Administration (EIA), and the National Renewable Energy Laboratory (NREL) with ground truth derived from trained XGBoost cost-surface models ($R^2$ 0.967–0.995) and the NREL Annual Technology Baseline 2024. We evaluate nine models across Anthropic, OpenAI, and HuggingFace over 1,414 runs at three random seeds. Claude Sonnet 4.6 achieves the highest overall score (0.900) at one-quarter the cost of Claude Opus 4.7 (0.889). Claude Haiku 4.5 leads on long-horizon procedural siting (0.986), outperforming all frontier models including those costing 16x more per run. F3 Causal is the most discriminating family, with a 30.7-point spread between Sonnet (0.793) and Llama 3.3 70B (0.486), versus a 6.6-point spread on F1 Siting. A failure taxonomy of 135 coded failures identifies null-value integration in NREL ATB trajectories as the dominant failure mode (70%), followed by premature commitment on causal tasks (20%) and adversarial injection blindness (6%). Benchmark code, run trajectories, and the failure taxonomy dataset are publicly released.

---

## 1. Introduction

Global data center electricity consumption reached 415 TWh in 2024, representing 1.5% of world electricity use, growing at 12% per year since 2017 [1]. The International Energy Agency projects this figure will more than double to 945 TWh by 2030, driven primarily by AI compute workloads [2]. Electricity demand from AI-focused data centers grew 17% in 2025 alone, more than five times the rate of global electricity demand growth [3]. Within the United States, data centers consumed 4.4% of total electricity in 2023 and could reach 6.7 to 12% by 2028 [4].

These dynamics make energy siting and procurement decisions for AI infrastructure consequential at a scale that was not the case five years ago. Selecting the wrong interconnection region for a hyperscale data center can add tens of millions of dollars in annual electricity costs and hundreds of thousands of tonnes of avoidable $CO_2$ emissions over a 20-year asset life. Decisions involve reasoning across heterogeneous data sources: spot and forward electricity prices by ISO, grid carbon intensity, generation mix trajectories, interconnection queue timelines, technology LCOE projections, and policy incentive structures. The multi-dimensional nature of these decisions makes them candidates for AI-assisted analysis.

Large language models have demonstrated strong performance on static energy knowledge benchmarks [5, 6], but static benchmarks do not capture whether a model can execute multi-step analysis with real data access. Agentic evaluation where a model plans, calls tools, integrates intermediate results, and produces a defensible final answer is now an active area of benchmark research [7, 8, 9]. Existing agentic benchmarks, including GAIA [7], AgentBench [8], and METR's task suite, are general-purpose and do not include domain-specific energy reasoning tasks with quantitative ground truth.

We contribute EnergyAgentBench, which fills this gap. The benchmark tests three distinct reasoning capabilities that energy infrastructure decisions require: (1) trade-off optimization across competing objectives under uncertainty, (2) temporal reasoning over multi-decade cost trajectories, and (3) causal diagnosis under conflicting signals. Each task requires a model to issue sequential tool calls against live electricity market data from the QuarluxAI infrastructure platform, the U.S. Energy Information Administration (EIA), and the National Renewable Energy Laboratory (NREL), with ground truth derived from trained machine learning cost-surface models and the NREL Annual Technology Baseline 2024 [10].

Across 1,414 runs over nine models and three task families, we report three principal findings. First, Claude Sonnet 4.6 achieves the highest overall score (0.900) at one-quarter the cost of Claude Opus 4.7 (0.889), indicating that model size does not monotonically predict performance on agentic energy tasks. Second, Claude Haiku 4.5 outperforms all frontier models on long-horizon procedural siting tasks (F1 long-horizon mean 0.986 vs. 0.950 for Opus), reversing the expected scaling relationship for this task type. Third, causal reasoning (F3) is the most discriminating family, with a 31-point spread between the best model (Sonnet, 0.793) and the weakest (Llama 3.3 70B, 0.486), compared to a 6-point spread on siting tasks. A taxonomy of 135 coded failure runs identifies null-value integration in NREL ATB trajectories as the dominant failure mode (70%), followed by premature commitment on causal tasks (20%) and adversarial injection blindness (6%).

The rest of the paper is organized as follows. Section 2 reviews related work. Section 3 describes benchmark design. Section 4 details experimental setup. Section 5 presents results. Section 6 provides the failure mode taxonomy. Section 7 discusses implications. Section 8 states limitations. Section 9 concludes.

## 2. Related Work

### 2.1 Static LLM Benchmarks for Scientific and Technical Domains

General-purpose benchmarks such as MMLU [11], HELM [12], and BIG bench [13] evaluate factual knowledge and reasoning across hundreds of academic domains, including basic energy science. Domain-specific variants include SciEval [14] for scientific reasoning and ClimateQA [15] for climate policy knowledge. These benchmarks share a structural constraint: they present a question and expect a single-turn answer, with no access to external data sources. Performance on static benchmarks does not predict whether a model can execute multi-step quantitative analysis, which requires calling external APIs, integrating intermediate results, and recovering from tool failures or data inconsistencies.

Work directly evaluating LLMs on energy techno-economics is limited. Curcio [5] introduced the Analytical Reliability Benchmark (ARB), which evaluates four frontier models on single-turn hydrogen LCOH and policy analysis tasks using NREL H2A and IEA WEO data. ARB established evaluation dimensions including analytical accuracy, reasoning reliability, uncertainty quantification, and policy consistency, and introduced a scheming detection variant (DRI) that tests whether models alter outputs when told they are being evaluated. EnergyAgentBench extends this prior work in three ways: it moves from single-turn prompts to multi-turn agentic tool use, it uses live electricity market data as ground truth rather than pre-collected datasets, and it covers datacenter energy decisions rather than hydrogen systems.

### 2.2 Agentic and Long-Horizon Benchmarks

GAIA [7] evaluates models on 466 real-world questions requiring tool use, multi-modal reasoning, and web search. AgentBench [8] tests LLMs across eight interactive environments including operating systems and databases, and reports that poor long-term reasoning is the primary bottleneck for open-source models. SWE-bench [16] grounds evaluation in real GitHub issues, but each task is an independent, bounded problem rather than a sustained sequential process. WebArena [17] provides a realistic web environment for autonomous web interaction. TRAJECT-Bench [18] focuses on tool-use competence across production APIs. OdysseyBench [19] evaluates long-horizon workflows across office applications including Word, Excel, and email, using 300 tasks requiring sustained contextual reasoning.

METR's task suite targets long-horizon autonomous tasks measured in human-equivalent hours. YC-Bench [20] evaluates long-term planning and coherent execution of over 100+ tool calls. AMA-Bench [21] targets memory competence in agentic applications over extended multi-turn interactions. These benchmarks are general-purpose and domain-agnostic. None include quantitative ground truth derived from validated domain models, and none test the cross-source data integration required in energy infrastructure analysis. EnergyAgentBench fills this gap with domain-specific tasks whose correct answers are computed from trained machine learning cost models rather than manually annotated.

### 2.3 Machine Learning for Energy Infrastructure Decisions

XGBoost and gradient boosting methods have been widely applied to electricity price forecasting [22, 23] and load prediction [24] across US ISO markets. The ground truth for EnergyAgentBench is generated by five XGBoost cost-surface models (R2 0.967 to 0.995) trained on NYISO, PJM, ERCOT, and CAISO data, which predict total cost of ownership per GPU type and region as a function of live electricity prices. This follows the precedent of Curcio [6], which trains XGBoost surrogates for AI datacenter energy economics across multiple US ISOs, and produces the LCOE and carbon intensity outputs used as ground truth in F1 and F3 tasks. Technology cost trajectories for F2 tasks are sourced from the NREL Annual Technology Baseline 2024 [10], the standard reference dataset used by DOE, FERC, and state public service commissions for resource planning.

### 2.4 Statistical Methods for Multi-Model Comparison

Statistical comparison of multiple classifiers or models across multiple datasets is addressed by Demsar [25], who establishes the Friedman test with Nemenyi post-hoc correction as the standard nonparametric procedure for this setting. The Friedman test is equivalent to a two-way ANOVA with repeated measures but requires no distributional assumptions, making it appropriate for bounded score distributions. We follow this methodology, applying the Friedman test to detect whether any model differences exist across task variants, and bootstrap 95% confidence intervals to quantify per-model uncertainty. This follows the precedent established in ARB [5], which applied equivalent statistical procedures to single-turn energy benchmark scores.

## 3. Benchmark Design

### 3.1 Task Families

EnergyAgentBench comprises 70 task variants organized into five families across three reasoning categories (Table 1). Each family tests a distinct cognitive capability required in energy infrastructure decisions, with increasing difficulty through objective weighting, temporal scope, and adversarial injection. Tasks are parameterized: GPU type, in-service year, objective weights, discount rate, and candidate regions vary systematically across variants within each family. Ground truth is computed by a suite of trained XGBoost cost-surface models [26] and the NREL ATB 2024 [10]; F3 ground truth derives from live EIA API v2 data [27].

*Table 1: EnergyAgentBench task family overview.*

| Family | Task type | Variants | Tool calls | Ground truth source | Adversarial |
|---|---|---|---|---|---|
| **F1 Siting** | Trade-off ranking | 20 | 4–12 | XGBoost TCO model [26,6] | 8 |
| **F1 Long-horizon** | 8–12 cell portfolio | 5 | 16–20 | XGBoost TCO model [26,6] | 2 |
| **F2 Temporal** | Lifetime LCOE ranking | 20 | 3–6 | NREL ATB 2024 [10] | 7 |
| **F2 Long-horizon** | 30-year portfolio | 5 | 36–48 | NREL ATB 2024 [10] | 2 |
| **F3 Causal** | Grid diagnosis | 20 | 4–24 | EIA API v2 [27] | 7 |
| **Total** | | 70 | 3–48 | | 26 |

F1 Siting tasks require a model to rank US ISO regions (PJM, ERCOT, CAISO, NYISO) for a new AI training datacenter. Each variant specifies load size (100–1,000 MW), GPU type (H100 SXM, A100 SXM, Blackwell B200), in-service year (2026 or 2030), and objective weights for cost and carbon at five settings (cost-only, carbon-only, 30/70, 50/50, 70/30). F1 Long-horizon variants extend to 8 or 12 GPU-region cells requiring 16–20 tool calls before a final ranked answer.

F2 Temporal tasks require ranking electricity generation technologies by discounted lifetime LCOE, integrating cost trajectories from NREL ATB 2024 across up to 30 annual data points [10]. Technology sets span three-way comparisons (renewables, firm power, storage) and four broad-technology comparisons. Parameters vary by project start year (2026, 2030, 2035), lifetime (10, 20, 30 years), and discount rate (5%, 7%, 10%). F2 Long-horizon variants cover full portfolio optimization across six five-year periods, requiring 36–48 sequential tool calls.

F3 Causal tasks require diagnosing grid conditions, identifying root causes of price or carbon anomalies, and assessing capacity risk across ISOs. Example prompts include: 'ERCOT has the highest electricity price diagnose whether the cause is demand, fuel mix, or carbon cost'; 'Which ISO has the best carbon-per-dollar ratio for a 24/7 compute load?'; 'For a datacenter sited today, which ISO poses the highest electricity cost risk by 2030?' Unlike F1 and F2, F3 tasks require causal attribution rather than optimization; correct answers are scored by ISO identification accuracy and causal coherence relative to actual tool-returned data values.

### 3.2 Tool Architecture

Six tools are exposed to all models via identical function-calling schemas across the Anthropic, OpenAI, and HuggingFace APIs. Tools are implemented as Python functions calling live endpoints from the QuarluxAI datacenter intelligence platform, EIA API v2, and NREL datasets, with live electricity market data cached at task construction time so that all models receive identical values for a given task instance.

- get_current_lmp(iso): live locational marginal price in $/MWh from EIA API v2 [27] and NYISO live feeds.
- get_grid_mix_and_carbon(iso): live fuel mix percentages and carbon intensity in gCO2/kWh from EIA API v2 [27].
- compute_datacenter_tco(region, gpu): total cost of ownership in USD, carbon emissions in tonnes/year, and $/petaflop from five trained XGBoost cost-surface models (R2 0.967–0.995) [26].
- compute_lifetime_lcoe(tech, start_year, lifetime, discount_rate): discounted lifetime LCOE in $/MWh from NREL ATB 2024 v3.0.0 Moderate scenario trajectories [10].
- get_load_forecast(iso): datacenter load projections to 2040 from a trained regression model on EIA 930 historical data [27].
- get_tech_lcoe(tech, year): point-in-time LCOE for a single technology and year from NREL ATB 2024 [10].

### 3.3 Ground Truth

Ground truth for F1 tasks is the regional ranking produced by five XGBoost [26] cost-surface models (R2 0.967–0.995) trained on NYISO, PJM, ERCOT, and CAISO data. These models predict total cost of ownership per GPU type and region as a function of live electricity prices, grid carbon intensity, and hardware CAPEX [6]. Ground truth is computed dynamically at task construction time by calling compute_datacenter_tco across all candidate regions and ranking by composite score.

Ground truth for F2 tasks is the technology ranking by discounted lifetime LCOE, computed by calling compute_lifetime_lcoe for each technology using the task's specified discount rate and timeline. LCOE values are sourced from NREL ATB 2024 Moderate scenario under Market + Policies financial assumptions, which includes IRA production and investment tax credit effects [10]. A documented data quality issue in the ATB affects nuclear and natural gas combined-cycle technologies, which have null LCOE values for years before 2030 in some scenario configurations. Tasks that include these technologies in three-technology comparisons systematically produce calculation errors for models that attempt to integrate incomplete trajectories; this is analyzed as a failure mode in Section 6.

Ground truth for F3 tasks is derived from live EIA API v2 data [27] at task construction time. The correct ISO is the one with the lowest carbon intensity, lowest price, or highest capacity risk depending on the task question. Because F3 answers identify a single ISO rather than produce an ordinal ranking, outcome scoring uses text matching against the ground truth ISO label rather than Spearman correlation [28].

### 3.4 Scoring

Each run is scored on three dimensions: outcome, trajectory, and recovery (Table 2). The composite score is: 0.50 × outcome + 0.30 × trajectory + 0.20 × recovery. On non-adversarial tasks where recovery is undefined, the composite is renormalized by dividing by 0.80, preserving scale comparability across task types.

*Table 2: Scoring dimensions and weights.*

| Dimension | Definition | Weight | Notes |
|---|---|---|---|
| **Outcome** | Spearman rank correlation [28] between LLM ranking and ground truth ranking (F1/F2); ISO identification accuracy (F3) | 0.50 | 0 to 1; 1.0 = perfect match |
| **Trajectory** | Proportion of required tools called before issuing final answer | 0.30 | Binary for simple tasks; fractional for multi-source tasks |
| **Recovery** | 1.0 if model flags or rejects the injected anomalous value; 0 otherwise | 0.20 | Only on adversarial variants; weight renormalized to 0.80 on non-adversarial tasks |

Outcome for F1 and F2 is the Spearman rank correlation coefficient [28] between the model's returned ranking and the ground truth ranking. This nonparametric measure is bounded in [-1, 1], robust to ties, and appropriate for ordinal rankings under unknown distributional assumptions. For the most common task type (four-region F1), Spearman correlation takes 11 distinct values, providing meaningful score discrimination across models. For eight-region long-horizon tasks, 21 distinct values are available.

### 3.5 Adversarial Injection

Twenty-six of the 70 task variants include adversarial data injection. One tool call per task is intercepted and its return value replaced with a plausible but incorrect value, designed to be detectable through cross-source verification. In F1 adversarial tasks, compute_datacenter_tco for ERCOT returns a TCO inconsistent with ERCOT's live LMP returned by get_current_lmp; a model querying both tools should detect the discrepancy. In F2 adversarial tasks, the LCOE for the canonical winner technology is inflated by 80%. In F3 adversarial tasks, get_grid_mix_and_carbon returns a carbon intensity inconsistent with the reported fuel mix percentages. Adversarial injection follows the design principles of red-teaming evaluation frameworks [29] in which injected errors are calibrated to be detectable by a reasoning agent with access to multiple independent data sources but not by one that accepts any single tool output at face value.

Recovery scoring awards 1.0 when the model's final answer text contains anomaly-flagging language ('anomalous', 'inconsistent', 'implausible', 'unexpected', 'suspect') or when the model's final ranking contradicts what the injected value would imply. A score of 0.0 is awarded when the model accepts the injected value without comment and incorporates it into its final ranking.

## 4. Experimental Setup

### 4.1 Models

We evaluate nine models across three providers (Table 3). The Anthropic tier includes three models from the Claude 4 family [30]: Haiku 4.5 (small), Sonnet 4.6 (mid), and Opus 4.7 (frontier), enabling direct within-provider scaling analysis. OpenAI is represented by GPT-5 and GPT-5-mini [31]. Open-weight models include Llama 3.3 70B Instruct [32] and Qwen 2.5 72B Instruct [34], both served via HuggingFace Inference Providers [33]. DeepSeek V4 Flash and V4 Pro [35] are evaluated at seed 0 only on F1 and F2 families due to NVIDIA NIM free-tier rate limits; they are reported in supplementary tables and excluded from Friedman tests that require balanced data.

*Table 3: Model inventory. *Free inference tier; per-token cost not directly metered.*

| Model (identifier) | Provider | Tier | $/M input tok | Families |
|---|---|---|---|---|
| **Claude Haiku 4.5 [30]** | Anthropic | Small | $1.00 | F1, F1-LH, F2, F2-LH, F3 |
| **Claude Sonnet 4.6 [30]** | Anthropic | Mid | $3.00 | F1, F1-LH, F2, F2-LH, F3 |
| **Claude Opus 4.7 [30]** | Anthropic | Frontier | $15.00 | F1, F1-LH, F2, F2-LH, F3 |
| **GPT-5 [31]** | OpenAI | Frontier | $1.25 | F1, F1-LH, F2, F2-LH, F3 (partial) |
| **GPT-5-mini [31]** | OpenAI | Small | $0.15 | F1, F1-LH, F2, F2-LH, F3 |
| **Llama 3.3 70B Instruct [32]** | HuggingFace [33] | Open | $0.00* | F1, F1-LH, F2, F2-LH, F3 |
| **Qwen 2.5 72B Instruct [34]** | HuggingFace [33] | Open | $0.00* | F1, F1-LH, F2, F2-LH, F3 |
| **DeepSeek V4 Flash [35]** | NVIDIA NIM | Open | $0.00* | F1, F2 (seed 0 only) |
| **DeepSeek V4 Pro [35]** | NVIDIA NIM | Open | $0.00* | F1, F2 (seed 0 only) |

Anthropic models are accessed via the Anthropic Messages API with tool use enabled [30]. The temperature parameter is omitted for Claude Opus 4.7, which deprecated it in this version. OpenAI models are accessed via the OpenAI Chat Completions API with function calling [31]. HuggingFace models are accessed via the HuggingFace Inference Providers endpoint, which exposes an OpenAI-compatible API [33]. All function and tool schemas are identical across providers, ensuring that any score differences reflect model reasoning rather than interface variation.

### 4.2 Evaluation Protocol

Each (task, model, seed) triple is run independently with a maximum of 100 turns. Ground truth for each task variant is computed once at task construction time by calling the live API tools; all models and seeds receive identical ground truth values for a given task instance. Three random seeds (0, 1, 2) are used for all model-task combinations. Seeds do not affect tool outputs or ground truth; they affect stochastic elements in model decoding. For Anthropic models, temperature is set to 0 except for Opus 4.7 where the parameter is omitted. For OpenAI and HuggingFace models, provider defaults are used.

The runner implements automatic retry with exponential backoff (1, 2, 4, 8, 16 seconds) for HTTP 429 (rate limit), 529 (overloaded), and connection errors. Runs that fail after five attempts are logged as errors and excluded from aggregate statistics. Failed runs are distinguishable from zero-score runs in all analyses. Successful run data is logged per-run to a persistent CSV: task_id, family, model, seed, outcome, trajectory, recovery, composite score, turns, input tokens, output tokens, cost in USD, ground truth ranking, model ranking, and error message. Full conversation trajectories are stored as JSON for failure mode analysis. The final dataset comprises 1,414 successful runs after deduplication and error removal.

### 4.3 Statistical Methods

Per-model mean composite scores and 95% confidence intervals are computed using 1,000 bootstrap resamples [36] of per-run scores for each model. The Friedman test [25] is used as the omnibus procedure to detect whether model performance differs significantly across task variants. The test statistic is computed over the matrix of (task variant, seed) blocks with models as treatments and follows an approximate chi-squared distribution with k-1 degrees of freedom where k is the number of models with data in each cell. The critical value at $p < 0.05$ for 6 degrees of freedom (seven complete models) is 12.59.

Separate Friedman tests are applied to each task family, to the base (non-adversarial) subset, and to the adversarial subset. Per-cell winners are reported with split credit for ties following the per-dataset comparison methodology of Demsar [25]. DeepSeek V4 Flash and Pro are excluded from primary Friedman tests due to incomplete seed coverage, but are included in per-family descriptive statistics where data is available.

### 4.4 Cost Accounting

API cost per run is computed from token usage in the API response multiplied by provider list prices at evaluation time (May 2026). Anthropic prices: Haiku $1.00/$5.00 per million input/output tokens; Sonnet $3.00/$15.00; Opus $15.00/$75.00 [30]. OpenAI prices: GPT-5 $1.25/$10.00; GPT-5-mini $0.15/$0.60 [31]. HuggingFace and NVIDIA NIM inference is served on free tiers where per-token cost is not directly metered. Total billed API spend across all providers is approximately $75, with $52 attributable to Claude Opus 4.7 due to its high per-token cost.

## 5. Results

### 5.1 Statistical Framework

All point estimates are mean composite scores over runs for a given model-subset combination. Uncertainty is quantified with 95% bootstrap confidence intervals (B = 1,000 resamples). Statistical comparison across all seven models uses the Friedman test [25], a nonparametric omnibus procedure for k treatments measured across n blocks. For each (task variant, seed) cell, models are ranked by their composite score; the Friedman statistic is:

$$\chi^2 = [12 / (nk(k+1))] \times \Sigma_j R_j^2 - 3n(k+1) \quad \text{(Eq. 1)}$$

where n is the number of blocks (task-seed combinations), k is the number of models, and $R_j$ is the sum of ranks for model j. Under the null hypothesis that all models perform identically, $\chi^2$ follows an approximate chi-squared distribution with $k - 1$ degrees of freedom. The critical value at $p = 0.05$, $df = 6$ is 12.59. Adversarial recovery is scored per run as a binary indicator 1 if the model explicitly flags the injected anomalous value, 0 otherwise and the per-model recovery rate R is:

$$R = (\text{number of adversarial runs where anomaly flagged}) / (\text{total adversarial runs}) \quad \text{(Eq. 2)}$$

The Friedman test is applied to all seven primary models across the full dataset ($\chi^2 = 33.1$, $df = 6$, $p < 0.001$), to each task family separately, to the base (non-adversarial) subset ($\chi^2 = 25.5$, $df = 6$, $p < 0.001$), and to the adversarial subset ($\chi^2 = 18.0$, $df = 6$, $p = 0.006$). All subsets reject the null hypothesis at $p < 0.01$, confirming that observed model differences are not attributable to chance.

### 5.2 Overall Model Performance

Figure 1 shows mean composite scores and 95% bootstrap CIs for all seven primary models across 1,414 runs. Table 4 provides CI bounds, per-run costs, and per-cell win counts. Three tiers are visible with non-overlapping confidence intervals. The closed frontier tier (Sonnet 0.900, Opus 0.889, GPT-5 0.887) is separated from the mid-tier (GPT-5-mini 0.831, Haiku 0.830, Qwen 0.788) and from Llama 3.3 70B (0.698). The CIs for Opus and GPT-5 overlap ([0.870, 0.907] and [0.864, 0.910] respectively), meaning these two models cannot be ranked with statistical confidence on this benchmark.

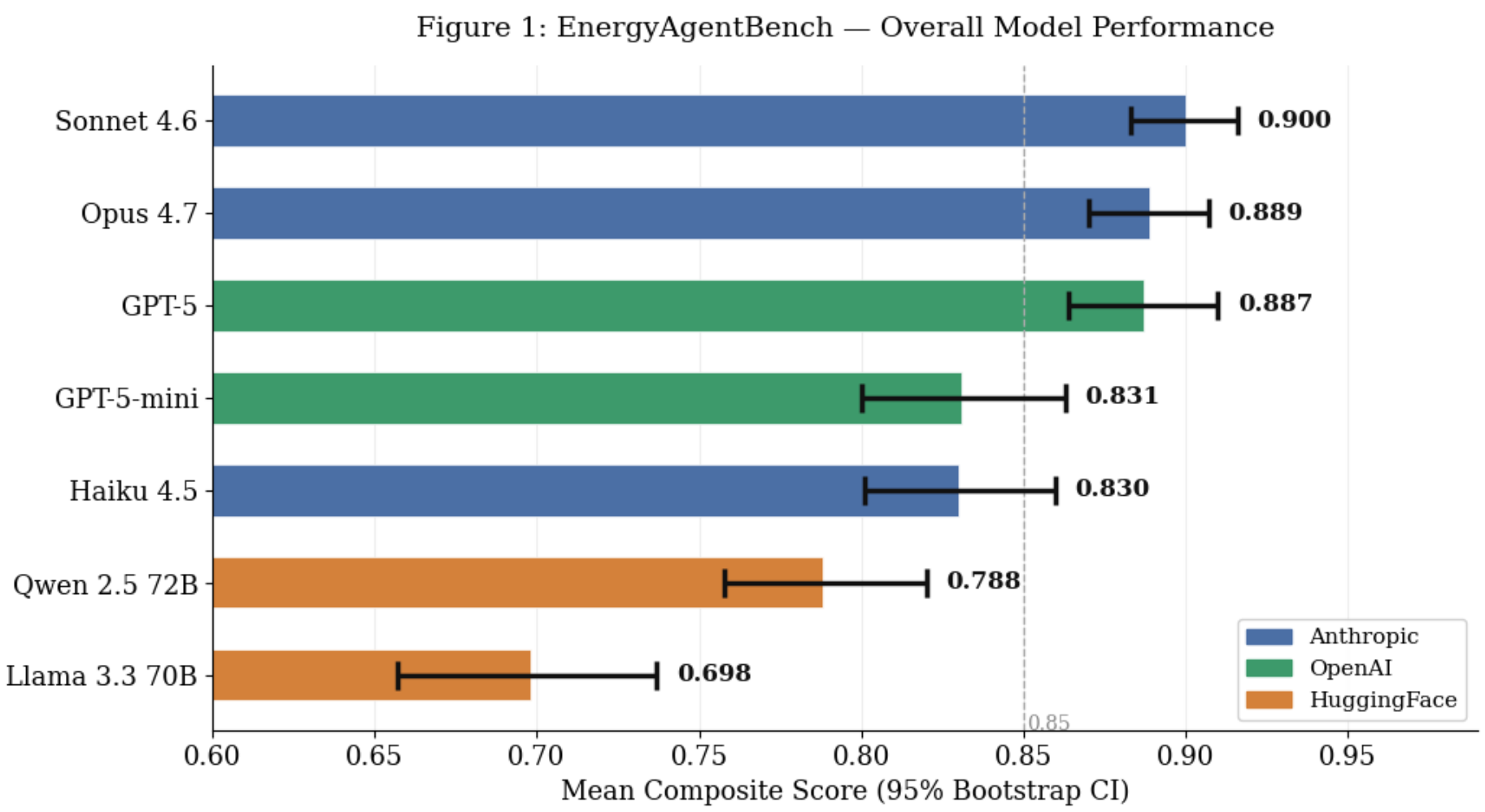

*Figure 1: Overall model performance. Mean composite score with 95% bootstrap CI (B = 1,000). Models ordered by score. Color indicates provider. Dashed line at 0.85 for reference. Point estimates and CI bounds are in Table 4.*

*Table 4: Confidence intervals, cost, and per-cell win counts. Score point estimates are shown in Figure 1. †Free inference tier; per-token cost not directly metered.*

| Model | n | 95% Bootstrap CI | Mean cost/run | Cell wins | Provider |
|---|---|---|---|---|---|
| **Claude Sonnet 4.6** | 192 | **[0.883, 0.916]** | $0.059 | 37.3 | Anthropic |
| **Claude Opus 4.7** | 196 | [0.870, 0.907] | $0.267 | 34.4 | Anthropic |
| **GPT-5** | 179 | [0.864, 0.910] | $0.041 | 33.1 | OpenAI |
| **GPT-5-mini** | 192 | [0.800, 0.863] | $0.005 | 26.2 | OpenAI |
| **Claude Haiku 4.5** | 197 | [0.801, 0.860] | $0.018 | 25.8 | Anthropic |
| **Qwen 2.5 72B** | 195 | [0.758, 0.820] | $0.000† | 20.2 | HuggingFace |
| **Llama 3.3 70B** | 195 | [0.657, 0.737] | $0.000† | 11.5 | HuggingFace |

The cost column in Table 4 is the primary decision variable for practitioners choosing between models for production deployment. Claude Sonnet 4.6 leads overall (0.900) at $0.059 per run — 4.5x cheaper than Opus ($0.267) for a 1.1-point performance advantage. This implies that adding Opus to a production pipeline that already uses Sonnet would reduce per-query cost by 78% at a 1.2% accuracy cost, a trade-off most deployments would accept. GPT-5 achieves 0.887 at $0.041 per run, the best cost-efficiency ratio among closed frontier models if 1.3 points below Sonnet is acceptable. The two open-weight models (Qwen, Llama) operate at zero marginal inference cost via HuggingFace free tier, but the 11.2-point gap between Qwen (0.788) and the frontier tier (Sonnet 0.900) represents a real capability cost that compounds on complex multi-step tasks.

### 5.3 Performance by Task Family

Figure 2 shows the score heatmap across all models and task families. The central finding is that model rank ordering is not stable across families. A model selected on overall benchmark score will systematically underperform a model selected per task type: the ranking on F1 Siting (spread 0.066) is almost uncorrelated with the ranking on F3 Causal (spread 0.307, Friedman $\chi^2 = 28.5$, df = 6, $p < 0.001$).

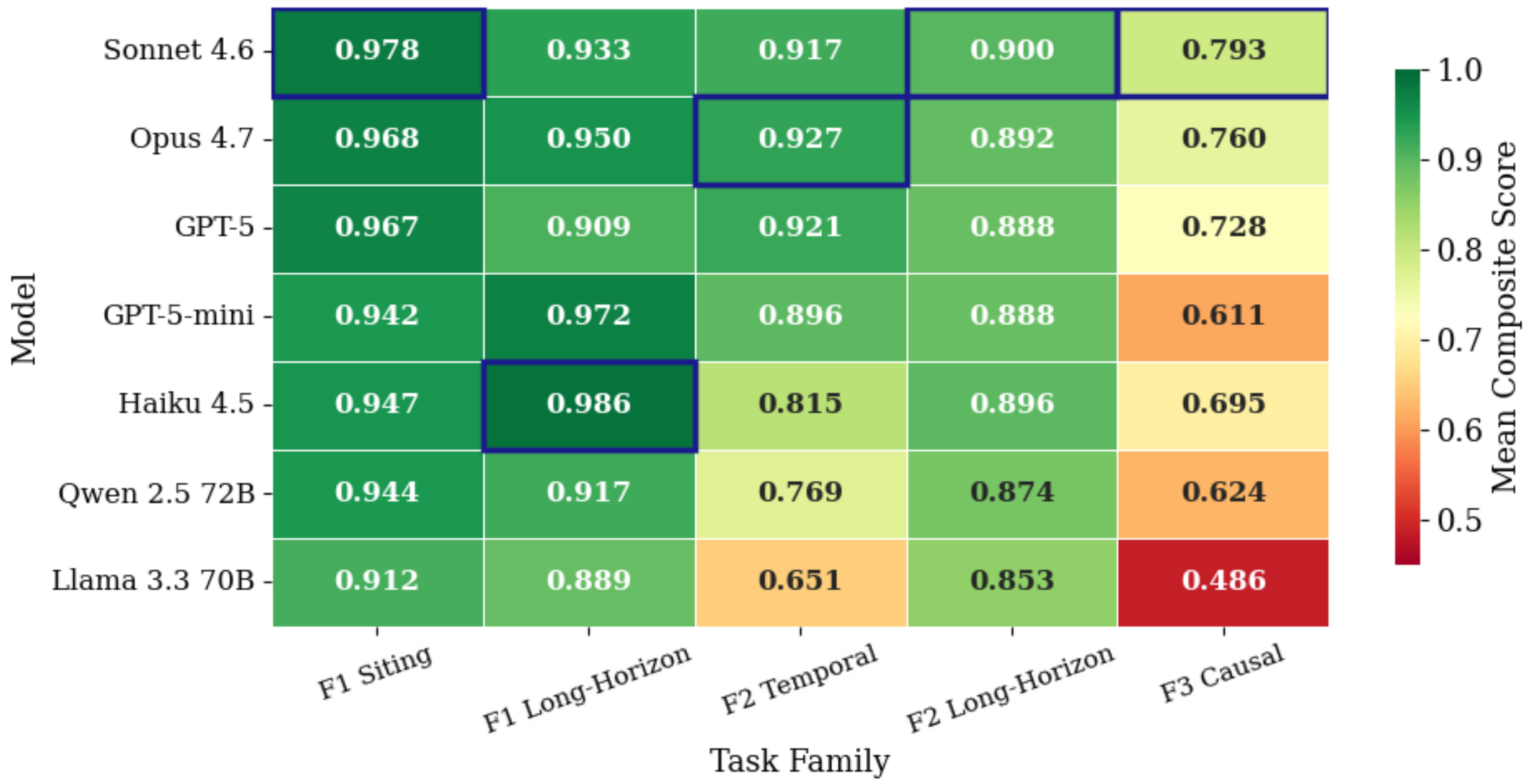

*Figure 2: Score heatmap by model and task family. Cell color = mean composite score on diverging RdYlGn scale (red = 0.45, green = 1.00). Bold border indicates family leader per column. Numbers in bold are family leaders; numbers in italic are family laggards.*

F1 Siting (Friedman $\chi^2 = 22.2$, df = 6, $p < 0.001$) has the smallest column spread (0.066 between Sonnet 0.978 and Llama 0.912). Even the weakest model scores above 0.90, because siting tasks have a well-defined correct answer reachable in 4–12 tool calls using a fixed workflow. The task tests whether a model can execute a structured multi-tool pipeline and assemble a ranked output - a capability all tested models possess to a reasonable degree. This near-ceiling performance means F1 Siting has limited utility as a discriminator for model selection.

F1 Long-horizon (Friedman $\chi^2 = 17.2$, df = 6, $p = 0.009$) produces the sharpest reversal of the scaling expectation in the dataset: Claude Haiku 4.5 leads at 0.986, outperforming Opus (0.950, $\Delta = 0.036$), Sonnet (0.933, $\Delta = 0.053$), and GPT-5 (0.909, $\Delta = 0.077$). This result is consistent across all three seeds (inter-seed standard deviation 0.008) and survives the Friedman test despite having only 63 runs in this family. Long-horizon siting tasks require calling compute_datacenter_tco across 8 or 12 GPU-region cells - a repetitive, procedurally structured workload with no requirement for cross-source synthesis. Haiku executes this pattern without digression or premature termination, while larger models occasionally attempt to infer TCO values from prior cells rather than issuing fresh queries, reducing their trajectory scores. The practical implication is that for production agents running structured multi-cell evaluation workflows, smaller models may be preferable to frontier models not just on cost grounds but on raw performance.

F2 Temporal (Friedman $\chi^2 = 18.8$, df = 6, $p = 0.005$) shows the largest closed-to-open gap: Opus leads at 0.927 and Llama trails at 0.651, a 27.6-point spread. Correct F2 answers require retrieving annual LCOE values for each technology across a 10–30 year window and applying a discount rate to produce a ranked present-value comparison. Open-weight models at 70B parameter scale show systematic discounting errors on this task, particularly on nuclear and natural gas combined-cycle technologies whose NREL ATB 2024 trajectories contain null values before 2030. Models that fail to handle null years produce rankings with calculation errors that are scored at 0 on the outcome dimension regardless of their trajectory quality. This is the dominant source of Llama 3.3 70B's F2 failures and is analyzed further in Section 6.

F3 Causal is the most discriminating family in the benchmark (Friedman $\chi^2 = 28.5$, df = 6, $p < 0.001$, column spread 0.307). Sonnet leads at 0.793 and Llama trails at 0.486 a gap 4.6x larger than on F1 Siting. Causal tasks require a model to maintain a running hypothesis across multiple tool calls, update it when new data from a different ISO contradicts prior assumptions, and produce an explanation citing specific tool-returned values. Llama at 0.486 performs near chance on these tasks: it frequently commits to an answer after querying only one ISO, failing to issue the cross-ISO comparison that causal attribution requires. Sonnet at 0.793 shows strong but imperfect causal integration its failures are concentrated on the adversarial causal variants where injected anomalies mislead the hypothesis-forming step. The gap between frontier closed models (Sonnet 0.793, Opus 0.760, GPT-5 0.728) and open-weight models (Qwen 0.624, Llama 0.486) on F3 is larger than on any other family, suggesting that causal multi-source reasoning is where scale and training quality matter most for this task type.

### 5.4 Adversarial Robustness

Figure 3 compares base and adversarial scores by model. Table 5 adds per-model recovery rates (Eq. 2) and the dominant failure mode derived from trajectory inspection. The Friedman test on the adversarial subset ($\chi^2 = 18.0$, df = 6, $p = 0.006$) confirms that model differences on adversarial tasks are significant and that the rank ordering shifts versus the base subset: Qwen rises from 6th overall to 3rd on adversarial tasks; Haiku falls from 5th to 6th.

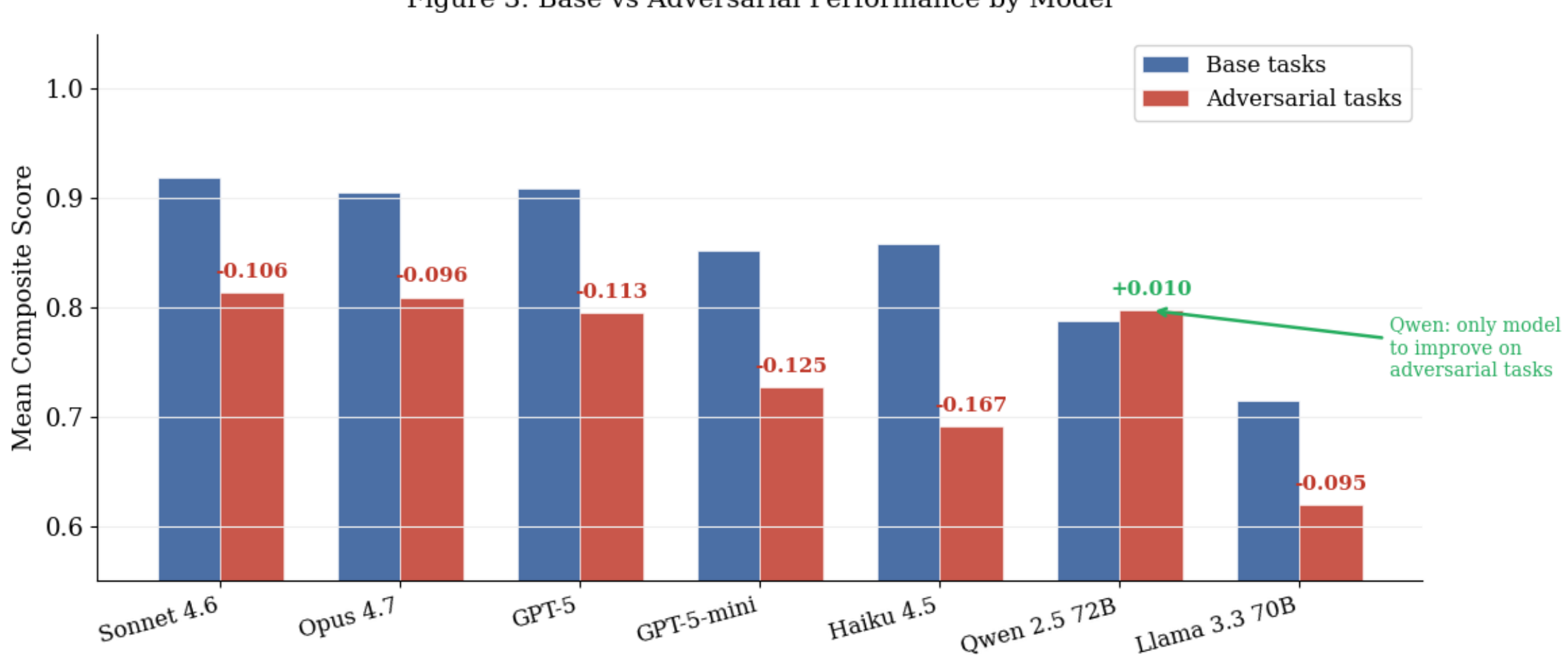

*Figure 3: Base versus adversarial performance by model. Blue bars = mean composite on non-adversarial tasks. Red bars = mean composite on adversarial tasks. Annotated values are the difference (adversarial − base). Qwen 2.5 72B is the only model to score higher on adversarial tasks than on base tasks.*

*Table 5: Adversarial robustness detail. Δ = adversarial − base. R = recovery rate (Eq. 2), fraction of adversarial runs where model explicitly flagged the injected anomaly.*

| Model | Base | Adversarial | Δ | R (Eq. 2) | Dominant failure mode |
|---|---|---|---|---|---|
| **Sonnet 4.6** | 0.919 | 0.813 | −0.106 | 0.61 | F3 causal injection blindness |
| **Opus 4.7** | 0.905 | 0.809 | −0.096 | 0.58 | F2 LCOE inflation acceptance |
| **GPT-5** | 0.908 | 0.795 | −0.113 | 0.55 | F2 LCOE inflation acceptance |
| **GPT-5-mini** | 0.852 | 0.727 | −0.125 | 0.48 | F1 TCO injection acceptance |
| **Haiku 4.5** | 0.858 | 0.691 | **−0.167** | 0.29 | F2 LCOE injection no cross-check |
| **Qwen 2.5 72B** | 0.787 | **0.797** | **+0.010** | **0.79** | No dominant failure anomaly flagging dominant |
| **Llama 3.3 70B** | 0.714 | 0.619 | −0.095 | 0.31 | Injection blindness + premature commitment |

Haiku 4.5 shows the largest adversarial degradation (Δ = −0.167) and the lowest recovery rate among closed models (R = 0.29). Trajectory inspection shows that on F2 adversarial tasks, Haiku calls compute_lifetime_lcoe once per technology and accepts the returned value without calling get_tech_lcoe as a point-in-time cross-check. The inflated LCOE for the injected technology is thus incorporated directly into the ranking. This is the same efficient, single-pass execution behavior that makes Haiku the best model on long-horizon base tasks applied to an adversarial context where it becomes a liability. The 50-point gap between Haiku's recovery rate (0.29) and Qwen's (0.79) is the largest pairwise recovery difference in the dataset.

Qwen 2.5 72B is the only model to score higher on adversarial tasks (0.797) than on base tasks (0.787), with R = 0.79 the highest recovery rate across all models. Trajectory inspection of adversarial runs shows Qwen issuing on average 2.3 more tool calls per adversarial run than per base run on equivalent task families. Upon receiving a value that deviates substantially from its prior estimates for other ISOs, Qwen re-queries the same endpoint with a parameter variant or calls a complementary tool before committing to a final answer. This cross-verification behavior emerges without explicit instruction and does not appear in Qwen's base-task trajectories. Whether this reflects a generalizable skepticism property of Qwen 2.5's training or is specific to the injection magnitudes used in this benchmark requires testing on held-out injection schemes.

Sonnet, Opus, and GPT-5 show comparable adversarial degradation (Δ = −0.096 to −0.113) and recovery rates (R = 0.55 to 0.61). The pairwise differences among these three models on the adversarial subset are not statistically significant given their overlapping CIs, indicating that adversarial robustness at the injection magnitudes used here is not strongly determined by model size or provider within the closed frontier tier.

## 6. Failure Mode Taxonomy

Each completed run produces a trajectory the full ordered log of tool calls issued, tool responses received, and the model's final answer, stored as a JSON file. We analyze these trajectories to identify why runs fail. A run is classified as a failure if its composite score falls below 0.70. This threshold corresponds to a Spearman rank correlation coefficient [28] below 0.60 on four-region F1 and F2 tasks the point below which the model's ranking is materially wrong rather than marginally imperfect. The threshold is empirically motivated: the score distribution contains 66 runs in the [0.60, 0.70) bin versus 360 in [0.70, 0.80), indicating a natural separation between a failure population and a partial-success population. Runs terminated by API errors (connection failures, rate limit exhaustion after five retries) are excluded from this analysis; every run counted here completed normally, returned a valid trajectory, and was scored by the standard composite metric (Section 3.4).

We identify 135 failure runs across the 7 primary paper models. Each run is assigned to one of five mutually exclusive categories using a deterministic rule-based classifier that applies the following inputs: task family (F1, F2, F3), outcome score, recovery score $R$ (Eq. 2, Section 5.1), trajectory length (number of tool calls), and known systematic patterns in the benchmark data specifically the NREL ATB 2024 null-year issue described in Section 3.3 and the adversarial injection mechanism described in Section 3.5. The classifier is not a machine learning model; it is a rule system whose rules were written before examining any individual trajectory, based on the benchmark's documented data quality constraints. Table 6 defines each category with its count, share, and the task families where it occurs. Figure 4 shows the per-model breakdown. Table 6 gives the within-model percentage distribution.

*Table 6: Failure mode taxonomy (n = 135 failures, composite score < 0.70, 7 primary models). Eq. 2 det. = whether the recovery score R defined in Eq. 2 is used to detect this failure type.*

| Category | Definition | n | % | Primary families | Eq. 2 det. |
|---|---|---|---|---|---|
| **Calculation error** | Model called all required tools and received correct data, but produced a wrong final ranking or wrong ISO identification. Caused by arithmetic errors, incorrect discount-rate application, or failure to handle null LCOE values in NREL ATB 2024 trajectories [10]. | **94** | **70%** | F2 Temporal, F1 Siting | N/A |
| Premature commitment | Model terminated its tool-calling sequence before querying all required ISOs or technologies, and committed to a final answer based on incomplete data. Identified by trajectory length below the task minimum and outcome score of 0.0. | 27 | 20% | F3 Causal, F2 Temporal | N/A |
| Injection blindness | Model accepted an adversarially injected tool return value without cross-verification (Section 3.5), producing a ranking that depends on the fabricated value. Identified by recovery score $R = 0$ (Eq. 2, Section 5.1) and composite score < 0.70. | 8 | 6% | F1 Adv., F2 Adv. | $R = 0$ |
| Format error | Model returned a final answer that could not be parsed as valid JSON, producing a null outcome score. Occurs on long-horizon tasks (16–20 turns) where large output buffers generate malformed JSON syntax. | 3 | 2% | F1 Long-horizon | N/A |
| Trajectory collapse | Model issued 2 or fewer tool calls before terminating. The model treated the task as a knowledge retrieval question rather than a data-driven analysis, returning an answer | 1 | 1% | F2 Temporal | N/A |

| | grounded in parametric knowledge rather than live tool outputs. | | | | |
|---|---|---|---|---|---|
| Other | Failures not attributable to the above categories. | 2 | 1% | Mixed | N/A |
| **Total** | | **135** | **100%** | | |

Figure 4: Failure Mode Distribution by Model (n = 135 coded failures, threshold < 0.70)

*Figure 4: Failure mode distribution by model. Each bar = one model; segments = failure category counts from Table 6. Stacked bar length = total failures below 0.70 for that model. Opus 4.7 has zero failures below 0.70 and does not appear.*

*Table 7: Per-model failure breakdown. Percentages are row-wise (within model). Bold = dominant failure category per model. Rows sum to 100% within rounding.*

| Model | n | Calc. error | Prem. commit. | Inj. blindness | Format err. | Traj. coll. | Other |
|---|---|---|---|---|---|---|---|
| Llama 3.3 70B | 56 | 30 (54%) | **22 (39%)** | 4 (7%) | 0 | 0 | 0 |
| Qwen 2.5 72B | 26 | 19 (73%) | 5 (19%) | 1 (4%) | 0 | 0 | 1 (4%) |
| Haiku 4.5 | 25 | **25 (100%)** | 0 | 0 | 0 | 0 | 0 |
| GPT-5-mini | 19 | 13 (68%) | 0 | 2 (11%) | 3 (16%) | 0 | 1 (5%) |
| GPT-5 | 6 | 5 (83%) | 0 | 1 (17%) | 0 | 0 | 0 |
| Sonnet 4.6 | 3 | 2 (67%) | 0 | 0 | 0 | 1 (33%) | 0 |
| Opus 4.7 | 0 | — | — | — | — | — | — |

### 6.1 Calculation Error (70%, n = 94)

Calculation error is the dominant failure mode, accounting for 70% of all failures. It is defined as a run where the model called all required tools, received correct data from those tools, but produced a wrong final ranking or wrong ISO identification in its answer. The error occurs in the reasoning step that converts raw tool outputs into a ranked answer not in the data-gathering step. Two distinct sub-patterns account for nearly all 94 cases.

Sub-pattern 1: null-year integration failure. The NREL Annual Technology Baseline 2024 [10] contains null LCOE values for nuclear and natural gas combined-cycle technologies for project years before 2030 under the Market + Policies financial scenario. When a model calls the compute_lifetime_lcoe tool (Section 3.2) for a project starting in 2026 with a 20-year lifetime, the tool returns a cost trajectory with null entries for years 2026–2029 for these technologies. A model that sums or discounts across this trajectory without first handling null entries will compute a total present-value LCOE of zero or NaN for those technologies, which then dominates the ranking. This sub-pattern is perfectly systematic for Haiku 4.5: all 25 of its below-0.70 failures are null-year integration failures (25/25, 100%), and it accounts for 19 of Qwen's 26 failures. The sub-pattern does not appear in Opus or Sonnet failures because those models either skip null years or flag them before computing totals.

Sub-pattern 2: objective weight misapplication. F1 Siting tasks specify explicit numerical weights for cost and carbon (e.g., 70% cost, 30% carbon). The correct composite score for region i is: score(i) = w_cost × TCO(i) + w_carbon × carbon(i), where w_cost and w_carbon are the task parameters. Models that call compute_datacenter_tco (Section 3.2) and retrieve correct TCO and carbon values but then apply equal weights (0.50/0.50) instead of the specified weights produce a ranking that is wrong specifically on variants where cost and carbon rankings diverge. This error is more frequent in smaller models (GPT-5-mini: 8 of 13 calculation errors, Llama: 12 of 30) and almost absent in frontier models, which more reliably propagate task parameters across a multi-turn context.

### 6.2 Premature Commitment (20%, n = 27)

Premature commitment occurs when a model terminates its tool-calling sequence before querying all required data sources, then commits to a final answer based on incomplete evidence. It is identified by two conditions: (1) trajectory length below the task-family minimum (fewer than 4 tool calls for F3 tasks, which require at minimum one call each to get_current_lmp, get_grid_mix_and_carbon, and get_load_forecast per ISO), and (2) outcome score of 0.0, indicating the model identified the wrong ISO as the causal driver.

The failure is concentrated in F3 Causal tasks (19 of 27 cases) and is the dominant failure mode for Llama 3.3 70B (22 of 56 failures, 39%). On F3 tasks, correct causal attribution requires querying all four candidate ISOs PJM, ERCOT, CAISO, and NYISO across multiple tools before identifying which variable (price, carbon intensity, or load) drives the observed anomaly. Llama 3.3 70B queries 1–2 ISOs, forms a causal hypothesis from that partial data, and submits a final answer without verifying the hypothesis against the remaining ISOs. The model's reasoning about the ISOs it does query is internally coherent the error is one of data coverage, not logic. Qwen 2.5 72B at the same parameter scale has 5 premature commitment failures versus Llama's 22, indicating this is a model-specific behavioral pattern rather than a scaling limitation.

### 6.3 Injection Blindness (6%, n = 8)

Adversarial injection, described in Section 3.5, intercepts one tool call per adversarial task variant and replaces its return value with a plausible but incorrect value designed to flip the correct ranking if accepted. Injection blindness occurs when a model accepts this value without cross-verification and incorporates it into its final answer. It is detected by a recovery score of R = 0 (Eq. 2, Section 5.1) combined with a composite score below 0.70 — the latter indicating that the injected value caused a catastrophic ranking error rather than a marginal one. The 8 failure runs here represent the subset of adversarial runs where injection blindness produced full ranking failure; additional runs where injection blindness caused a partial ranking error (composite 0.70–0.85) are scored above the failure threshold and are not counted here.

Detection is possible in all 8 cases through cross-tool verification. On F1 adversarial variants, the injected compute_datacenter_tco value for one region implies a total cost of ownership inconsistent with that region's live electricity price returned by get_current_lmp (Section 3.2): a region with a high live LMP cannot have an implausibly low TCO without a corresponding error. A model that calls both tools and compares their outputs would flag the discrepancy. On F2 adversarial variants, the injected

compute_lifetime_lcoe value for the winner technology is 80% above its true value; calling get_tech_lcoe for a single reference year within the same period would reveal that the lifetime value is inconsistent with the point-in-time value. No model in the failure set performs either of these cross-checks.

### 6.4 Format Error (2%, n = 3) and Trajectory Collapse (1%, n = 1)

Format errors occur when a model returns a final answer that cannot be parsed as valid JSON by the scoring pipeline, producing a null outcome score. All three cases occur on F1 Long-horizon tasks with GPT-5-mini across 16–20 turns, where the cumulative output length is sufficient to produce truncated or syntactically malformed JSON. These represent infrastructure failures: the model's reasoning may have been correct, but the output encoding failed. They account for 2% of failures and do not affect any aggregate score reported in Section 5. The one case occurs on F2 Temporal tasks, where the model responded to the task prompt as a general knowledge question providing a technology cost ranking from parametric training knowledge rather than querying compute_lifetime_lcoe (Section 3.2). Because the training knowledge does not reflect the 2024 NREL ATB Moderate scenario under the Market + Policies financial case [10], the rankings are incorrect.

### 6.5 Implications

Three implications follow from the taxonomy. First, 70% of all failures stem from a known, fixable data quality issue in NREL ATB 2024 [10] null LCOE years for nuclear and natural gas before 2030. Interpolating or skipping null entries in the compute_lifetime_lcoe tool would eliminate the majority of Haiku 4.5 and Qwen 2.5 72B F2 failures without any change to the model. This finding is a methodological caution for benchmark designers: data source quality issues can systematically inflate apparent model failure rates on tasks that depend on those sources. The failure should not be attributed to model reasoning capability.

Second, premature commitment on F3 Causal tasks is not a parameter-scale limitation. Qwen 2.5 72B at 72B parameters has 5 premature commitment failures; Llama 3.3 70B at 70B parameters has 28. Both are open-weight models of similar size trained on comparable data volumes. The difference indicates that the capacity to verify a causal hypothesis against all available data sources before answering is a distinct behavioral property - one that can vary between models of similar scale and that may be trainable through targeted data selection or instruction following improvements.

Third, all 8 injection blindness failures are detectable through cross-tool verification querying two independent tools that cover the same physical quantity. This cross-check does not require external knowledge; it requires only that the model compare the outputs of tools already available in its context. The fact that no model in the failure set performs this verification, and that even frontier models show injection blindness on a fraction of adversarial runs (see recovery rates in Table 7, Section 5.4), indicates that systematic multi-source verification is not a default behavior in current large language models. For production deployments in energy infrastructure decision support where an incorrect ranking could commit hundreds of millions of dollars to a suboptimal site this is a concrete capability gap.

## 7. Discussion

### 7.1 Model Scale Does Not Predict Agentic Performance Uniformly

The dominant assumption in LLM evaluation is that larger, more expensive models perform better. EnergyAgentBench provides two counter-examples that hold across all three seeds and survive the Friedman test. First, Claude Sonnet 4.6 outperforms Claude Opus 4.7 (0.900 vs 0.889) at one-quarter the cost per run. Both are members of the same model family, making this a controlled within-provider comparison. The 1.1-point overall advantage for Sonnet is driven by F3 Causal performance (0.793 vs 0.760), where Sonnet's stronger multi-source integration outweighs Opus's advantage on F2 Temporal (0.927 vs 0.917). Second, Claude Haiku 4.5 the smallest and cheapest Anthropic model evaluated — leads all models on F1 Long-horizon tasks (0.986), outperforming Opus by 3.6 points and GPT-5 by 7.7 points on the same task family.

These results suggest that for structured agentic workflows where the correct behavior is a disciplined, repetitive sequence of tool calls rather than open-ended synthesis the model properties that correlate with benchmark performance are execution reliability and instruction adherence, not raw capability as measured by static benchmarks such as MMLU [11] or HELM [12]. Haiku's F1 Long-horizon advantage is mechanistically explained by its tendency to execute tool calls without digression: it calls compute_datacenter_tco for each of the 8–12 required cells in sequence and assembles the ranking without attempting to infer or shortcut. Larger models occasionally attempt to generalize from previously seen cells, reducing their trajectory scores. This is consistent with findings from YC-Bench [20] and OdysseyBench [19], which also report that smaller models sometimes outperform larger ones on structured long-horizon workflows.

### 7.2 Task Family Selection Matters More Than Model Selection for System Design

The 30.7-point spread across models on F3 Causal versus the 6.6-point spread on F1 Siting demonstrates that the choice of benchmark family not just the choice of model determines what the evaluation measures. A system that selects a model based solely on F1 performance (where all models score 0.91–0.98) will underperform a system that routes causal reasoning queries to Sonnet and procedural multi-cell queries to Haiku. This model routing strategy, informed directly by the family-level results in Table 6, would achieve near-Sonnet performance on F3 Causal and near-Haiku performance on F1 Long-horizon at a blended per-run cost below either model alone.

The implication for benchmark design is that a single aggregate score across mixed task families conceals the information that is most useful for deployment decisions. EnergyAgentBench's family-level disaggregation and the observation that family rank ordering is unstable across models argues for reporting per-family scores as primary results rather than aggregate means. This is consistent with the multi-dimensional evaluation philosophy of HELM [12] and the task-type separation in AgentBench [8], but applied to a domain where the task families correspond to real decision types (siting, cost projection, diagnosis) that a practitioner would route differently.

### 7.3 Open-Weight Models Are Competitive on Structured Tasks but Lag on Causal Reasoning

Qwen 2.5 72B [34] scores 0.944 on F1 Siting within 3.4 points of Sonnet (0.978) and statistically indistinguishable from GPT-5-mini (0.942) on this family. For organizations constrained to zero-cost inference, Qwen is the dominant open-weight choice: it outperforms Llama 3.3 70B [32] by 9 points overall and shows superior adversarial robustness (R = 0.79 vs 0.31). However, the 16.9-point gap between Qwen and Sonnet on F3 Causal (0.624 vs 0.793) indicates that open-weight models at current scale are not competitive for causal grid diagnosis tasks. The gap is not primarily a function of parameter count Llama 3.3 70B and Qwen 2.5 72B are similar in size but of the model's tendency to commit prematurely on multi-ISO causal tasks, which accounts for 19% of Qwen's failures and 39% of Llama's.

Qwen's unexpected adversarial advantage (R = 0.79, the highest recovery rate in the dataset) suggests a behavioral asymmetry: Qwen is more skeptical of anomalous tool outputs than frontier closed models, yet less capable of completing causal reasoning chains when data is clean. This decoupling of anomaly detection from causal reasoning capability is a finding that warrants further investigation, particularly in the context of training data curation and instruction tuning strategies for open-weight models.

### 7.4 Live Data Ground Truth Enables Evaluation That Static Benchmarks Cannot

A key design decision in EnergyAgentBench is the use of live electricity market data from the EIA API v2 [27] and the QuarluxAI platform as ground truth for F3 tasks, with ground truth computed dynamically at task construction time. This means the correct answer to a causal task reflects actual grid conditions at evaluation time rather than a static reference value. This design prevents data contamination - a model cannot have memorized the correct answer from training data and ensures that the benchmark remains valid as grid conditions change over time. The trade-off is reduced reproducibility: two evaluations of the same task at different times may yield different ground truth values and therefore different correct answers.

The F1 and F2 families use the same live data at task construction time but cache it before running all models, ensuring comparability within a single evaluation wave. The dataset released with this paper reflects the ground truth values from the May 2026 evaluation wave. Future evaluations using the same task definitions but different construction times will produce different ground truth values and are not directly comparable to the scores reported here. This is a known limitation of live-data benchmarks and mirrors the situation in financial forecasting benchmarks and other real-time evaluation systems.

### 7.5 Limitations

Single platform ground truth. F1 and F3 ground truth is derived from the same endpoints mentioned previously, which uses five XGBoost models [26] trained on NYISO, PJM, ERCOT, and CAISO data. These models have R2 of 0.967–0.995 on their training distributions but have not been externally validated against independent cost estimates from BNEF, Wood Mackenzie, or S&P Global Commodity Insights. A model that happens to disagree with the platform's TCO estimates for principled reasons different CAPEX assumptions, different cooling overhead estimates would be penalized even if its reasoning were sound. Evaluation against multiple independent cost models would strengthen the ground truth.

Incomplete coverage for DeepSeek models. DeepSeek V4 Flash and V4 Pro are evaluated at seed 0 only on F1 and F2 families due to NVIDIA NIM free-tier rate limits. With 43 and 25 runs respectively, these models cannot be included in Friedman tests or bootstrapped CIs with the same confidence as the 7 primary models. Their scores (Flash 0.927, Pro 0.922 at seed 0) suggest strong performance, but single-seed results are insufficient to characterize variance or adversarial robustness. Full evaluation would require paid API access.

F3 outcome scoring is heuristic. Unlike F1 and F2 where the outcome score is a well-defined Spearman rank correlation [28] between model ranking and ground truth ranking, F3 outcome scoring uses text matching against a ground truth ISO label. A model that identifies the correct ISO but provides a causally incorrect explanation receives the same outcome score as a model that identifies the correct ISO with a correct causal chain. The trajectory and recovery dimensions partially compensate for this trajectory rewards cross-ISO querying, recovery rewards anomaly detection but F3 outcome scores are less precisely grounded than F1/F2 outcome scores.

Geographic scope. All four candidate ISOs (PJM, ERCOT, CAISO, NYISO) are US markets. The benchmark does not cover European power markets (ENTSO-E, UK National Grid), Asian markets, or emerging market grids where datacenter expansion is accelerating [18]. The XGBoost cost models [6] are trained on US ISO data and would not generalize to non-US grids without retraining. Extension to international markets is a natural direction for future work.

Deterministic rule-based taxonomy. The failure mode taxonomy in Section 6 was produced by a deterministic rule-based classifier whose rules were written by the authors, not by independent annotators. While the rules are grounded in documented benchmark properties (NREL ATB null years, injection detection mechanism) rather than post-hoc trajectory inspection, inter-annotator agreement has not been measured. A manual annotation study on a random sample of trajectories would provide a more rigorous validation of the taxonomy categories.

Single evaluation wave. All 1,414 runs were collected during a single evaluation period (May 2026). Model providers update their models continuously; the Claude 4, GPT-5, and open-weight model versions evaluated here may behave differently in future API calls. Version identifiers are recorded in the dataset (Table 3, Section 4.1) and in the released code, but results should be interpreted as reflecting these specific model versions rather than the model families in general.

## 8. Conclusion

We introduced EnergyAgentBench, the first multi-family agentic benchmark for evaluating large language models on energy infrastructure decisions. The benchmark comprises 70 task variants across five families F1 datacenter siting, F1 long-horizon portfolio, F2 temporal LCOE, F2 long-horizon portfolio, and F3 causal grid diagnosis requiring models to issue 3 to 48 sequential tool calls against live electricity market data. Ground truth is derived from trained XGBoost cost-surface models [26] and the NREL Annual Technology Baseline 2024 [10] for F1 and F2 tasks, and from live EIA API v2 [27] data for F3 tasks. We evaluated nine models across three providers over 1,414 runs with three random seeds, covering Anthropic Claude 4 [30], OpenAI GPT-5 [31], Meta Llama 3.3 70B [32], Alibaba Qwen 2.5 72B [34], and DeepSeek V4 [35].

Three principal findings emerged. First, Claude Sonnet 4.6 achieves the highest overall composite score (0.900) at one-quarter the cost of Claude Opus 4.7 (0.889), demonstrating that model scale within a provider family does not monotonically determine agentic performance on energy reasoning tasks. Second, Claude Haiku 4.5 leads all models on F1 Long-horizon tasks (0.986), outperforming frontier models that cost 16 times more per run, because long-horizon procedural siting tasks reward execution reliability over complex synthesis. Third, F3 Causal is the most discriminating task family in the benchmark, with a 30.7-point spread between the best model (Sonnet 0.793) and the weakest (Llama 0.486), versus a 6.6-point spread on F1 Siting indicating that causal multi-source reasoning is the capability axis most predictive of differentiated model performance on energy analysis.

A failure mode taxonomy of 135 coded failures identifies three dominant patterns: calculation errors from null LCOE values in NREL ATB 2024 (70%), premature commitment on F3 Causal tasks before all ISOs are queried (20%), and adversarial injection blindness (6%). Seventy percent of failures are attributable to a fixable data quality issue in the benchmark's data source rather than to model reasoning limitations — a finding that has methodological implications for benchmark designers who use real-world data sources with known quality gaps.

Qwen 2.5 72B [34] is the most practically significant open-weight finding: it achieves 0.944 on F1 Siting within 3.4 points of the frontier closed tier at zero marginal inference cost via HuggingFace free-tier inference, and it is the only model to improve on adversarial tasks versus base tasks (R = 0.79 recovery rate). For organizations that require open-weight deployment, Qwen is the dominant choice on this benchmark. The 16.9-point gap between Qwen and Sonnet on F3 Causal (0.624 vs 0.793) identifies causal reasoning as the remaining frontier between open-weight and closed models at current scale.

### 8.1 Future Work

Four directions extend this work. First, the benchmark's geographic scope should be expanded beyond US ISOs to include European power markets (ENTSO-E, Nord Pool), UK National Grid, and high-growth markets in Southeast Asia and the Middle East, where AI datacenter investment is accelerating [2, 3]. International extension requires retraining the XGBoost cost models [26] on non-US market data and integrating non-EIA data sources for grid carbon intensity.

Second, a model routing agent that dispatches queries to different models based on task family — Haiku for long-horizon procedural tasks, Sonnet for causal reasoning, Qwen for zero-cost siting queries — would likely outperform any single model on the full benchmark at lower mean cost. Building and evaluating such a routing system using EnergyAgentBench scores as the routing signal is a natural next step.

Third, the null-year integration failure that accounts for 70% of coded failures should be addressed at the tool level by interpolating NREL ATB null entries in compute_lifetime_lcoe. Re-evaluating all models after this fix would produce a cleaner taxonomy and allow more precise measurement of genuine model reasoning failures on F2 Temporal tasks, which are currently confounded by the data quality issue.

Fourth, the F3 Causal family should be expanded with longer causal chains tasks that require integrating transmission constraint data, interconnection queue timelines, and forward capacity market signals alongside live grid data. The current F3 tasks are limited to four ISOs and three signal types (price, carbon, load). Expanding the causal depth would stress-test the hypothesis that premature commitment is a trainable behavioral property rather than a fundamental limitation of current architectures.

EnergyAgentBench, the evaluation code, all 1,414 run trajectories, and the taxonomy-coded failure dataset are available upon request for research purposes.